# Atomic force microscopy study of the tetragonal to monoclinic transformation behaviour of silica doped yttria-stabilized zirconia

Sylvain Deville, Jérôme Chevalier [#], Laurent Gremillard [a]

National Institute of Applied Science, Materials Department, Associate Research Unit 5510 (GEMPPM), 20 av. A. Einstein, 69621 Villeurbanne, France.
[a] current address: Materials Science Division, LNBL, Berkeley CA 94720, USA

[#] corresponding author. Tel: +33 4 72 43 61 25, Fax: +33 4 72 43 85 28.
E-mail: jerome.chevalier@insa-lyon.fr

The tetragonal to monoclinic phase transformation of zirconia has been the subject of extensive studies over the last 20 years [1-4]. The main features of the transformation have been identified and its martensitic nature is now widely recognised [5-8]. More specifically, the relevance of a nucleation and growth model to describe the transformation is widely accepted. Recent fracture episodes [9] of zirconia hip joint heads were reported, failures related to the t-m transformation degradation. Among the materials solutions considered for decreasing the sensitivity to t-m phase transformation, the possibility of adding silica as a dopant appears as an appealing one. Previous studies have revealed the beneficial effect of silica addition by the formation of a glassy phase at the grain boundaries and triple points. This glassy phase has been proven to reduce the residual stresses level [10], slowing down the transformation kinetics. Preliminary quantitative investigations by XRD have shown these materials are less susceptible to transformation. However, the mechanism by which the transformation propagated has still to be assessed.

Among the methods used to investigate this transformation [11-14], scanning light interferometry (SLI) was used [11-13] to quantify the growth rate of monoclinic spots. Both diameter and height were found to vary linearly with ageing time. If the vertical resolution is satisfying, the lateral resolution does not allow measuring the spots behaviour for diameter less than 2.5 μm in the work of Chevalier et al. and 5 μm in that of Grant et al. The very first stages of the transformation, of prime importance for subsequent growth behaviour, are not accessible. The possibility of using atomic force microscopy (AFM) to observe monoclinic spots growth with an improved precision was first reported by Tsubakino [15] and then Deville et al. [8]. This paper aims at showing AFM might be used to asses the nucleation and growth nature of the transformation for this material and describes quantitatively and very precisely the monoclinic spots growth rate from the very first stages of growth, providing thus much more reliable informations than SLI or traditional methods such as X-rays diffraction (XRD).

A high-purity powder (Tosoh TZ3Y-S with 3 mol% $Y_2O_3$ content) was used as starting powder. The materials were manufactured by a slip-casting method (slurries with 80 wt% solids content). 2.5 wt % of colloidal silica (Ludox HS40, Aldrich Chemical Company) was added to the slurries. The slurries were placed in plaster moulds in order to eliminate water and to form green compacts, which were then dried at 298 K in air for 7 days. Organic compounds were removed by heating at 873 K (heating rate: 15 K/h). Compacts were then sintered at 1723 K for 5 h in air (heating and cooling rate: 300 K/h). Samples were polished with standard diamond based products. AFM experiments were carried out with a D3100 nanoscope (Digital Instruments Inc.). The vertical resolution of AFM allows one to follow very precisely the transformation-induced relief. Ageing treatments were conducted in autoclave at 413 K, in water vapour atmosphere, in order to induce the phase transformation at the surface of the samples with time.

AFM height images are shown in Fig. 1. In the first image, after 5 min at 413 K, the transformation of three grains is clearly observed. The shear and volume increase accompanying the t-m transformation leads to a local uplift of the surface. This surface uplift is easily detected by AFM, and the transformed zones appear as brighter in AFM height images, where the contrast is proportional to the relief. Theses monoclinic spots will later on grow in diameter and height as seen on the following images, as the ageing treatment time increases, while new spots also appear elsewhere at the surface. It is worth noticing the growth of the spots can be followed very precisely, since AFM offers a very high lateral resolution, in the nanometer range, to be compared with the 2 micrometers lateral resolution of optical interferometry [13].The very first stages of spots growth can be therefore observed and precisely quantified. It is quite clear from these observations that the transformation is occurring by a nucleation and growth mechanism.

This aspect was quantitatively investigated, by following the growth of about 20 monoclinic spots. It is worth noticing that transformed zones resulting from the merging of two monoclinic spots was not taken into account, since the parameter of interest is the growth of a single spot and not of a spots assembly.

Considering the spots were approximately round shaped, the equivalent radius r of the spots was simply calculated from the surface by:

$$r = \sqrt{\frac{A}{\pi}}$$

where A is the measured monoclinic spot area. The surface and radius values are plotted in Fig. 2. The equivalent radius is clearly proportional to the spot age. A constant diameter growth rate of 5.8 $\mu m.h^{-1}$ is measured from the plot.

The spots height evolution was also measured as a function of their age, and is plotted in Fig. 3. A very fast increase is observed during the first 5 minutes. This increase is related to the growth of the first transformed martensitic variants [8]. Once these variants are formed, the transformation of the surrounding parts of the surface is triggered by the transformation induced stresses, reaching the same height as the first transformed variants. Further evolution of the spot height will then be governed by an eventual propagation of the transformation into the bulk.

Using the image relief contrast provided by AFM, image analysis was performed to quantify the fraction of transformed surface. Results are plotted in Fig. 4, and reveal the typical shape for the curve of an underlying nucleation and growth mechanism, in agreement with Fig. 1 observations. The monoclinic phase fraction was also measured by XRD, using the Garvie and Nicholson method [16]. Results are plotted on the same figure for comparison. XRD is much less sensitive to the surface degradation, one of the reasons being that the signal comes not only from the surface but also from a certain penetration depth of X-Rays. By comparing the curves in Fig. 4 and considering the monoclinic spots height variations in Fig. 3, it can be concluded that for this particular material, the transformation is propagating quite rapidly at the surface, but very slowly into the bulk of the material. This particular behaviour may therefore be related to the presence of the glassy phase at grain boundaries, which reduce the internal stresses and accommodates some of the transformation strain, reducing therefore the amount of microcracking that allows the water penetrating into the bulk of the samples, promoting the transformation of previously unexposed tetragonal grains.

AFM was used to characterize and quantify the ageing behaviour parameters at the surface of a silica doped yttria stabilised zirconia. Transformation occurred by a nucleation and growth mechanism. The monoclinic spots diameter was found to be directly proportional to ageing time, at a growth rate of 5.8 $\mu m.h^{-1}$. The presence of a glassy phase at the grain boundaries and triple points reduced the level of internal residual stresses, slowing down the propagation of the transformation into the bulk.

AFM appeared as an extremely powerful method to investigate the transformation at its first stages, with an improved resolution compared to XRD, as it was found to be much more sensitive to monoclinic phase fraction variations.

**Acknowledgements**

Authors are grateful to the CLAMS for using the nanoscope. Financial support of the European Union (GROWTH2000, project BIOKER, reference GRD2-2000-25039) is acknowledged.

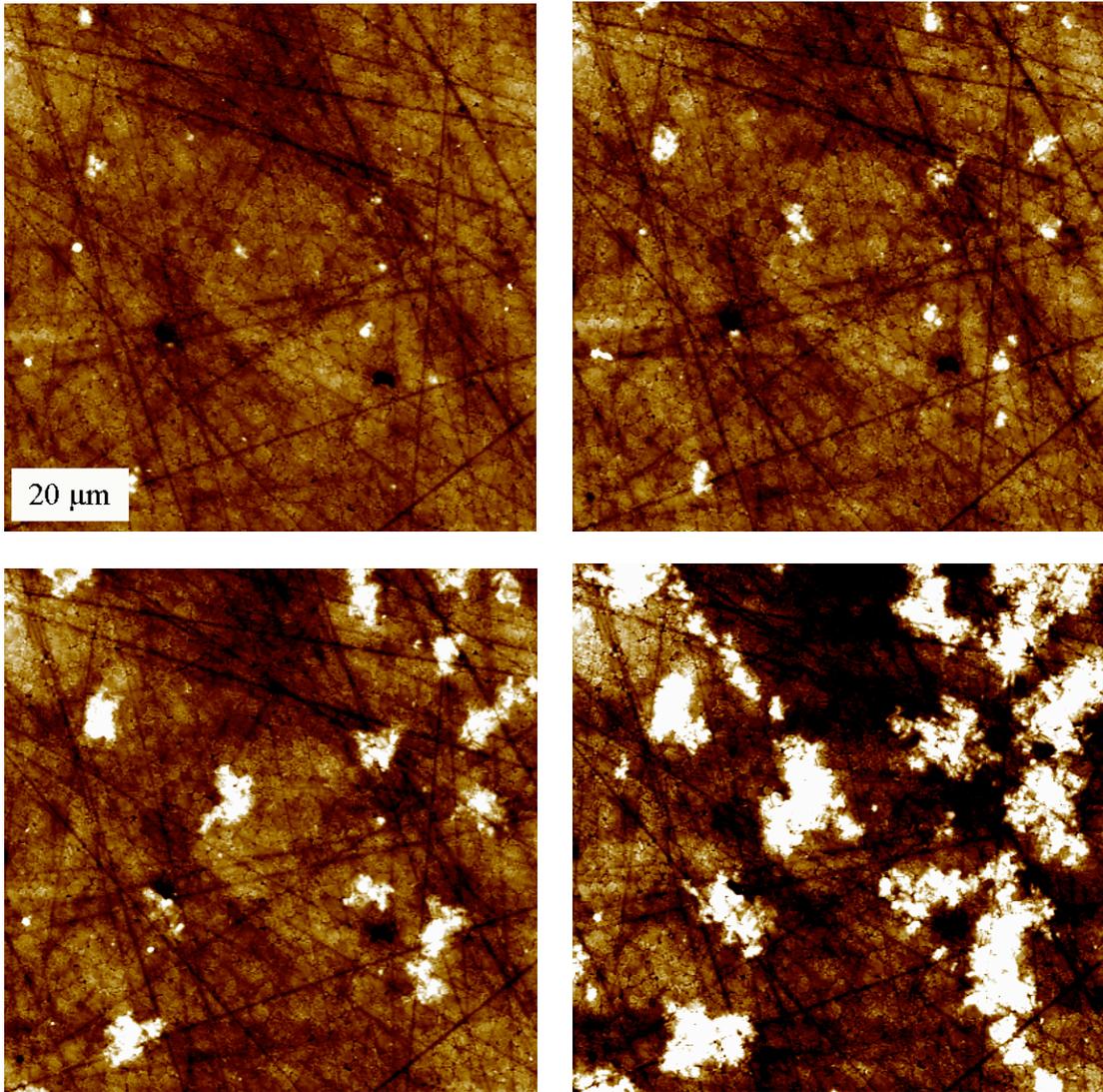

Fig 1: AFM observation of the transformation evolution of the same zone of the sample, at several times of the ageing treatment (5, 10, 20 and 35 min). Residual polishing scratches (about 5 nm deep) could be observed (dark straight lines).

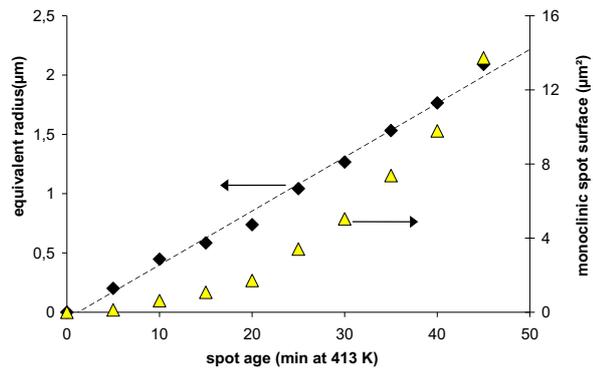

Fig. 2: Monoclinic spots average surface (∆) and equivalent radius (♦) as a function of their age.

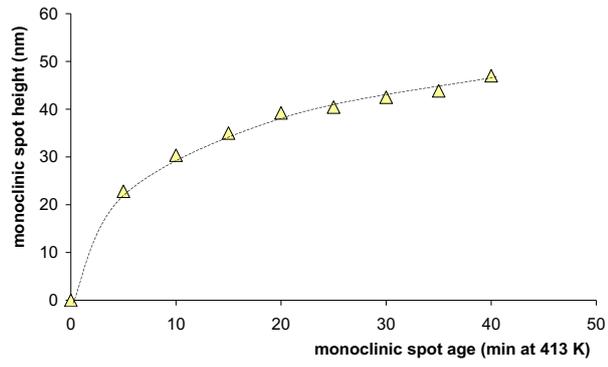

Fig. 3: Monoclinic spots height variation, as a function of their ages.

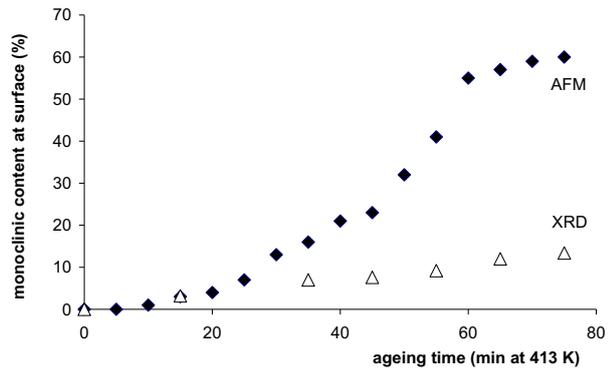

Fig. 4: Evolution of the transformed fraction with ageing time, measured by AFM (♦) and XRD (Δ).